\documentclass[
prd,superscriptaddress,eqsecnum,nofootinbib,notitlepage,
twocolumn
]{revtex4-1}

\usepackage[utf8]{inputenc}
\usepackage{hyperref}
\usepackage{graphicx}
\usepackage{amsmath,amssymb}
\usepackage{wasysym}

\newcommand{\be}{\begin{equation}}
\newcommand{\ee}{\end{equation}}
\newcommand{\ba}{\begin{eqnarray}}
\newcommand{\ea}{\end{eqnarray}}
\newcommand{\MSb}{$\overline{\text{MS}}$}

\newcommand{\GeV}{\text{\,GeV}}

\begin{document}

\title{Why should we care about the top quark Yukawa coupling?$^1$}
\footnotetext[1]{To be published in a special edition of the Journal of Experimental and Theoretical Physics in honor of the 60th birthday of Valery Rubakov.}
\setcounter{footnote}{1}

\author{Fedor Bezrukov}
\email{fedor.bezrukov@uconn.edu}
\affiliation{CERN, CH-1211 Gen\`eve 23, Switzerland}
\affiliation{Physics Department, University of Connecticut, Storrs, CT 06269-3046, USA}
\affiliation{RIKEN-BNL Research Center, Brookhaven National Laboratory, Upton, NY 11973, USA}
\author{Mikhail Shaposhnikov}
\email{mikhail.shaposhnikov@epfl.ch}
\affiliation{ Institut de Th\'{e}orie des Ph\'{e}nom\`{e}nes Physiques, \'{E}cole Polytechnique F\'{e}d\'{e}rale de Lausanne, CH-1015 Lausanne, Switzerland}
\date{\today}

\begin{abstract}
In the cosmological context, for the Standard  Model to be valid up to the scale of inflation, the top quark Yukawa coupling $y_t$ should not exceed the critical value  $y_t^{crit}$, coinciding with good precision (about 0.2{\text\permil}) with the requirement of the stability of the electroweak vacuum. So, the exact measurements of  $y_t$ may give an insight on the possible existence and the energy scale of new physics above 100~GeV, which is extremely sensitive to $y_t$. We overview the most recent  theoretical computations of  $y_t^{crit}$ and the experimental measurements of $y_t$. Within the theoretical and experimental uncertainties in $y_t$ the required scale of new physics varies from $10^7$~GeV to the Planck scale, urging for precise determination of the top quark Yukawa coupling.
\end{abstract}

\maketitle

\section{Introduction}
In the Spring of 2014 Valery Rubakov was visiting CERN and joined a bunch of theorists for a lunch at the CERN canteen.  As often happens, the conversation turned to the future of high energy physics: what kind of questions should be answered and what kind of experiments  should be done. Valery was arguing for the high energy frontier which would allow to search for new physics, whereas the authors of this article  brought attention to the precision measurements of  the top quark Yukawa coupling.  We remember Valery asking: ``Why should we care about the top quark Yukawa coupling?''  Because of some reasons the interesting  discussion was interrupted and we did not have a chance to explain our point of view in detail.  We use this opportunity to congratulate Valery with his coming jubilee and give an answer to his question in writing.  We apologise to Valery for describing in this text a number of well-known to him facts, which we included to make this essay accessible to a wider audience.

\section{Standard Model and the scale of new physics}
After the discovery of the Higgs boson at the LHC the Standard Model (SM) became a complete theory in the sense that all the particle degrees of freedom that it contains theoretically have been found experimentally. Moreover there are no convincing deviations from the SM in any type of high energy particle physics experiments. This raises a number of questions: ``Have we got at last the ultimate theory of Nature?'' and ``If not, where we should search for new physics?''

The answer to the first question is  well known and it is negative. The reasons are coming from the observations of  neutrino oscillations, absent in the SM, and from cosmology---the SM cannot accommodate dark matter and baryon asymmetry of the Universe. The last but not the least is the inflation, or, to stay strictly on the experimental evidence side, the flatness and homogeneity of the Universe at large scales and the origin of the initial density perturbations.  On a more theoretical side, the list of the drawbacks of the SM is quite long and includes incorporation of gravity into a quantum theory,  the hierarchy problem, the strong CP-problem, the flavour problem, etc., etc.  

The answer to the second question is not known. What is theoretically clear, is that some type of new physics must appear near the Planck energies $M_P=2.435 \times 10^{18}$ GeV, where gravity becomes important, but these energies are too high to be probed by any  experimental facility. The naturalness arguments put the scale of new physics close to the scale of electroweak symmetry breaking (see, e.g.~\cite{Giudice:2008bi, Giudice:2013yca}), but it is important to note that the SM by itself is a consistent quantum field theory up to the very high energies exceeding the Planck mass by many orders of magnitude, where it eventually breaks down due to the presence of Landau-poles in the scalar self-interaction and in U(1) gauge coupling.

As for the experimental evidence in favour of new physics, it does not give any idea of its scale: the neutrino oscillations can be explained by addition of Majorana leptons with the masses ranging from a fraction of electron-volt to $10^{16}$~GeV,  the mass of particle candidates for dark matter discussed in the literature vary by at least 30 orders of magnitude, the mass of Inflaton can be anywhere from hundreds of MeV to the GUT scale, whereas the masses of new particles responsible for baryogenesis can be as small as few MeV and as large as the Planck scale.  

As we are going to argue in this paper at the \emph{present moment} the \emph{only quantity} which can help us to get an idea about the scale of new physics is the top Yukawa coupling $y_t$. It may happen that the situation will change in the future: the signals of new physics may appear at the second stage of the LHC, or the lepton number violation will be discovered, or anomalous magnetic moment of muon will convincingly be out of the SM prediction, or something unexpected will show up.

\section{Vacuum stability and cosmology}
In the absence of  beyond the SM (BSM) signals the only way to address the question of the scale of new physics is to define the energy where the SM becomes theoretically inconsistent or contradicts some observations.  Since the SM is a renormalizable quantum field theory, the problems can appear only because of the renormalization evolution of some coupling constants, i.e.\ when they become large (and the model enters strong coupling at that scale), or additional minima of the effective potential develop changing the vacuum structure.  The most dangerous constant\footnote{The only other problematic parameter is the U(1) hypercharge which develops Landau pole, but only at the energy scale significantly exceeding Planck mass.} turns out to be the Higgs boson self-coupling constant $\lambda$ with the RG evolution at one loop
\begin{multline*}
  16\pi^2\frac{d\lambda}{d\ln\mu} =
  24\lambda^2+12\lambda y_t^2 -9\lambda(g^2+\frac{1}{3}g'^2) \\
  -6y_t^4 
  +\frac{9}{8}g^4+\frac{3}{8}g'^4
  +\frac{3}{4}g^2g'^2.
\end{multline*} 
The right hand side depends on the interplay between the positive contributions of the bosons and negative contribution from the top quark.
Before the discovery of the Higgs it was customary to show the results as a function of the Higgs mass $M_h\simeq\sqrt{2\lambda(\mu=M_h)}v$, with other parameters of the SM fixed by experiments.  For the Higgs mass $M_h>175$~GeV  the Landau pole in the Higgs self-coupling constant $\lambda$ occurs at energies smaller than the Planck scale, and comes closer to the Fermi scale when the mass of the Higgs boson is increasing \cite{Maiani:1977cg,Cabibbo:1979ay,Lindner:1985uk}.  For small Higgs masses the coupling becomes negative at some scale, and if the Higgs mass is below 113~GeV, the top quark loops give an essential contribution to the Higgs effective potential, making our  vacuum unstable with the life-time smaller than the age of the Universe \cite{Krasnikov:1978pu,Hung:1979dn,Politzer:1978ic}.\footnote{We should note that, strictly speaking, the Universe lifetime depends strongly on the form of the Planck scale suppressed higher-dimensional  operators in the effective action \cite{Branchina:2013jra}.}

The Higgs boson found at the LHC has a mass $M_h \simeq 125.7\pm0.4$~GeV \cite{Agashe:2014kda}  which is well within this interval. This means that the life-time of our vacuum exceeds that of the Universe by many orders of magnitude (see, e.g.~\cite{EliasMiro:2011aa}) and that  the SM without gravity is a weakly coupled theory even for energies exceeding the Planck scale also by many orders of magnitude. So, it looks that we cannot get any hint about the scale of new physics from these considerations. However, this is not true if we include in analysis the history of the Universe starting from inflation till the present time.

\begin{figure}[tbh]
  \centering
  \includegraphics{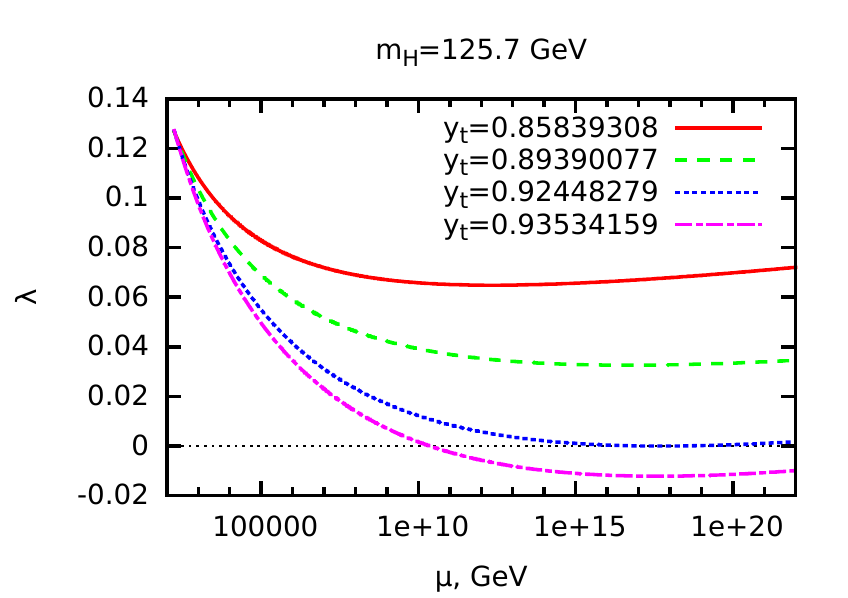}
  \caption{Renormalization group running of the Higgs coupling constant $\lambda$ for the Higgs mass $M_h=125.7$~GeV and several values of the top quark Yukawa $y_t(\mu=173.2\GeV)$.}
  \label{fig:lambda}
\end{figure}

\begin{figure}[tbh]
  \centering
  \includegraphics{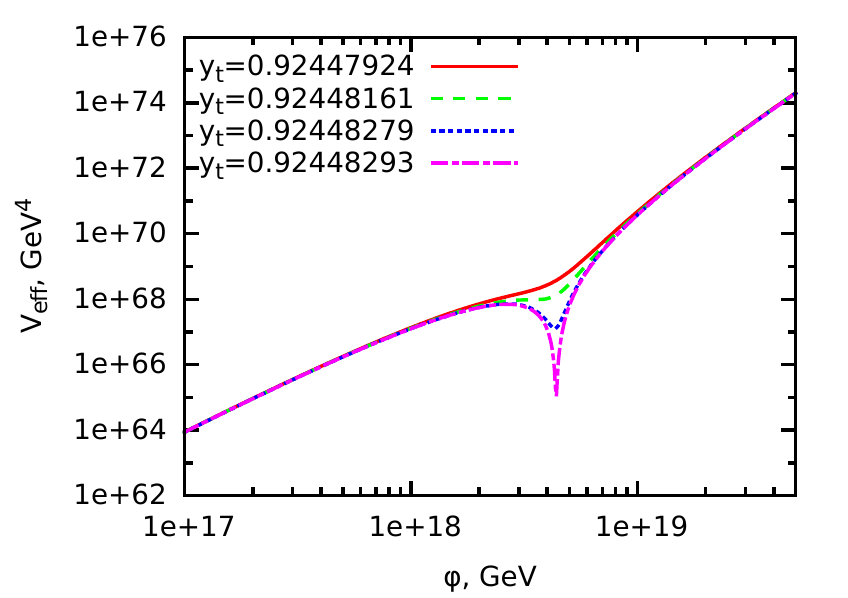}
  \caption{A very small change in the top Yukawa coupling  $y_t$ (taken at scale $\mu=173.2$~GeV) converts the monotonic behaviour of the effective potential for the Higgs field to that with an extra minimum at large values of the Higgs field.}
  \label{vefft}
\end{figure}

Since we want to get an idea about new physics, a way to proceed is to assume that there is none up to the Planck scale and see if we run to any contradiction. We can start from the SM without gravity and have a look at the effective potential for the Higgs field.  The contribution of the top quark to the effective potential is very important, as it has the largest Yukawa coupling to the Higgs boson. Moreover, it comes with the minus sign and is responsible for appearance of the extra minimum of the effective potential at large values of the Higgs field. We fix all parameters of the SM to their experimental values except the top Yukawa coupling (we will see below that at present it is the most uncertain one for the problem under consideration). For definiteness, we will be using the \MSb{} subtraction scheme  and take $y_t$ at some specific normalization point $\mu=173.2$~GeV.  Then, the RG evolution of the Higgs coupling $\lambda$ for various top quark Yukawas is illustrated by Fig.~\ref{fig:lambda}. Close to the ``critical'' value of the top Yukawa coupling, to be defined exactly right away, the effective potential (\ref{Veff}) behaves as shown in Fig.~\ref{vefft}. For $y_t<y_t^\text{crit}-1.2\times10^{-6}$ it increases while the Higgs field increases, for $y_t>y_t^\text{crit}-1.2\times10^{-6}$ a new minimum of the effective potential develops at large values of the Higgs field, at $y_t= y_t^\text{crit}$ our electroweak vacuum is degenerate with the new one, while at $y_t> y_t^\text{crit}$ the new minimum is deeper than ours, meaning that our vacuum is metastable. If $y_t>y_t^\text{crit}+0.04$ (this corresponds roughly to the top quark mass $m_t\gtrsim178$~GeV) the life-time of our vacuum is smaller than the age of the Universe.

The case $y_t<y_t^\text{crit}-1.2\times10^{-6}$ is certainly the most cosmologically safe, as our electroweak vacuum is unique. However, if $y_t>y_t^\text{crit}-1.2\times10^{-6}$ the evolution of the Universe should lead the system to our vacuum rather than to the vacuum with large Higgs field (as far as our vacuum is the global minimum). While in the interval $y_t \in (y_t^\text{crit}-1.2\times10^{-6}, y_t^\text{crit})$ our vacuum is deeper than another one so that the happy end is quite plausible, it is not so for $y_t> y_t^\text{crit}$, when it is the other way around.

\begin{figure*}
  \centering
  \includegraphics{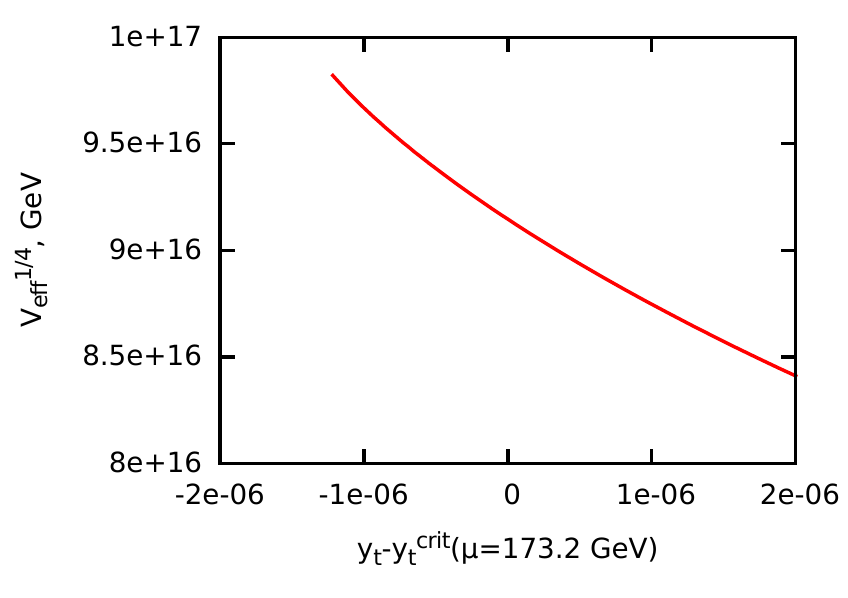}%
  \includegraphics{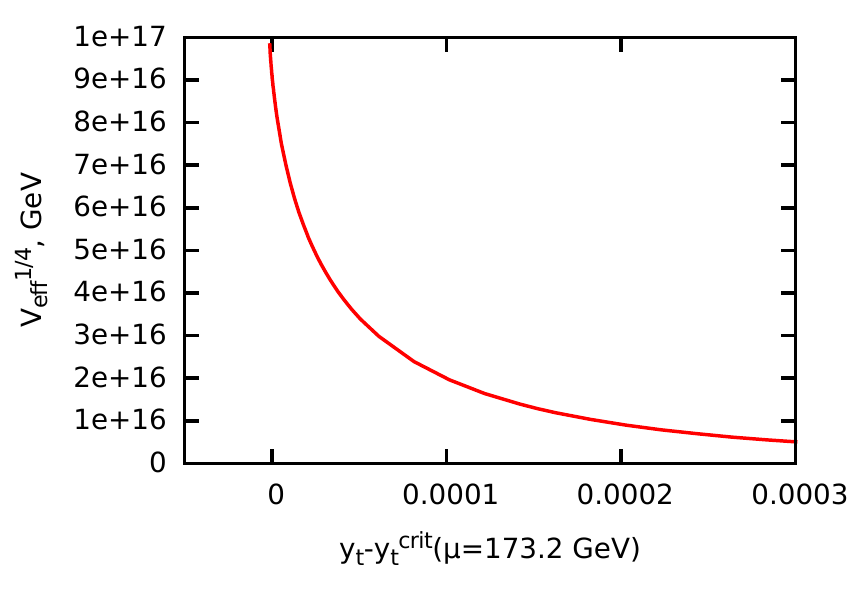}
  \caption{Height of the potential barrer near the critical value $y_t^{\text{crit}}$.}
  \label{fig:vmax}
\end{figure*}

In order to understand how far one can go from the (absolutely) safe values $y_t\leq y_t^\text{crit}$ into the dangerous region, we can consider yet another feature of the effective potential---the value of the potential barrier which separates our electroweak vacuum from that at large values of the Higgs field. The energy density corresponding to this extremum is gauge-invariant and does not depend on the renormalization scheme.  It is presented in Fig.~\ref{fig:vmax}. Now, if the Hubble scale at inflation does not exceed that of the potential barrier, it is conceivable to think that the presence of another vacuum is not important, while in the opposite situation the de-Sitter fluctuations of the Higgs field would drive the system to another vacuum.  And, indeed, several papers \cite{Kobakhidze:2013tn,Enqvist:2014bua} argued that this is exactly what is going to happen. 

Of course, this statement is only true if the potential for the Higgs field is not modified by the gravitational effects or by the presence of some new physics at the inflationary scale. For example, as has been shown in  \cite{Herranen:2014cua}, the addition of even a small non-minimal coupling $\xi<0,~|\xi| \sim 10^{-2}$ of the Higgs field $\phi$ to the Ricci scalar $R$, 
\be
\left(\frac{M_P^2}{2}  + \xi \phi^2\right) R
\label{nonmin}
\ee
increases the barrier height and thus stabilise the vacuum against fluctuations induced by inflation. Taken at the face value the action (\ref{nonmin}) with negative $\xi$ leads to instabilities at large values of the background Higgs field, but this can be corrected by considering a more general case, replacing  $\xi \phi^2$ by a function of the Higgs field that never exceeds $M_P^2/2$ \cite{Kamada:2014ufa}. At the same time, the presence of the non-minimal coupling with the opposite sign would severely destabilise the vacuum.

We do not know yet what was the energy density $V_\text{inf}$ at inflation, as this depends on the value $r$ of the tensor-to-scalar ratio as
\be
  V_\text{inf}^{1/4} \sim
  1.9\times10^{16}\GeV\left(\frac{r}{0.1}\right)^{1/4}~.
\ee
For the BICEP II  value of $r\simeq 0.2$ \cite{Ade:2014xna} this energy is $2.3\times10^{16}$~GeV.  Then the requirement discussed above leads to the constraint on the top Yukawa $y_t<y_t^\text{crit}+0.00009$, with the deviation from $y_t^\text{crit}$ being numerically very small. Because of a very weak dependence of $V_\text{inf}$ on $r$, even for Starobinsky $R^2$ inflation \cite{Starobinsky:1980te} or for non-critical Higgs inflation \cite{Bezrukov:2007ep}, which have a much smaller tensor-to-scalar ratio $r\simeq 0.003$, the resulting constraint is just a bit weaker, $y_t<y_t^\text{crit}+0.00022$. Let us denote this small positive deviation from $y_t^\text{crit}$ by $\delta y_t$, depending on $r$.

To summarise, if the measurement of top quark Yukawa will give us $y_t<y_t^\text{crit}+ \delta y_t$, the embedding of the SM without any kind of new physics  in cosmology does not lead us to any troubles and thus no information on the scale of new physics can be derived.  This would however be a great setting for the ``SM like'' theories without new particles with masses larger than the Fermi scale \cite{Asaka:2005pn,Asaka:2005an,Bezrukov:2007ep,Shaposhnikov:2006xi,Bezrukov:2013fca}.

Suppose now  that  $y_t>y_t^\text{crit}+\delta y_t$. In this case one can get some idea on the scale of new physics by the following argument (see, e.g. \cite{Degrassi:2012ry} and references therein). Let us consider the value of the scalar field at which the effective potential crosses zero (we normalise $V_\text{eff}$ in such a way that it is equal to zero in our vacuum). Or, almost the same, the normalization point $\mu_\text{new}$ where the scalar self-coupling $\lambda$ crosses zero, indicating an instability at this energies.\footnote{To be precise, the value of the scalar field  where the effective potential is equal to zero is gauge-noninvariant and depends of renormalization scheme. The value of $\mu$ where the scalar self-coupling constant crosses zero is scheme dependend but is gauge invariant, if the gauge-invariant definition of $\lambda$ is used, as in  \MSb{}. In what follows we will be using \MSb{} subtraction scheme and the effective potential in the Landau gauge. The use of other schemes or gauges can change $\mu_\text{new}$ by two orders of magnitude or so \cite{Loinaz:1997td,DiLuzio:2014bua,Andreassen:2014gha,Nielsen:2014spa}.} 

\begin{figure}
  \centering
  \includegraphics{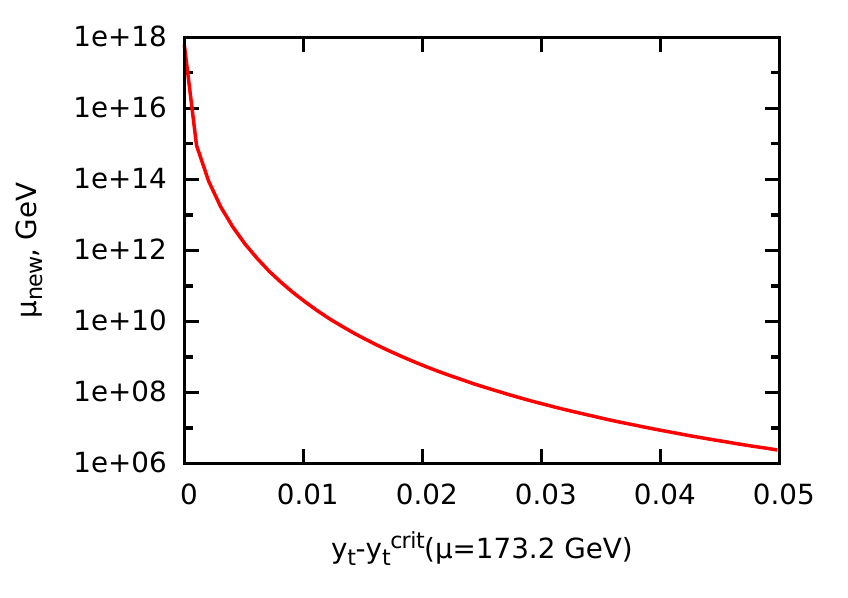}
  \caption{Scale $\mu_0$ where the Higgs self-coupling $\lambda$ becoming negative (possibly requiring new physics at lower energies) depending on the top quark Yukawa $y_t$.}
  \label{fig:munew}
\end{figure}

To make the potential or scalar self-coupling positive for  all energies, something new should intervene at the scale around or below $E\simeq \mu_\text{new}$.  There are many possibilities to do so, associated with existence of new thresholds, new scalars or fermions with masses $\lesssim \mu_\text{new}$ \cite{Chen:2012faa,EliasMiro:2012ay,Baek:2012uj,Allison:2012qn,Kanemura:2012ha,Lee:2013nv,Coriano:2014mpa,Belanger:2014bga}.  Fig.~\ref{fig:munew} shows the dependence of the scale $\mu_\text{new}$ on $y_t$. One can see that it is very sharp: in the vicinity of $y_t^\text{crit}$ the change of $y_t$ by a tiny amount  leads to a change in $\mu_\text{new}$ by many orders of magnitude!  Though exactly what kind of new physics would be needed remains to be an open question, these facts call for a precise experimental measurements of $y_t$. 

\section{Computation of the critical top Yukawa coupling}
\label{sec:comp}

To find the numerical value of $y_t^\text{crit}$,  one should compute the effective potential for the Higgs field $V(\phi)$ and determine the parameters  at which it has two degenerate minima:
\be
  V(\phi_{SM})=V(\phi_1),  \quad  V'(\phi_{SM})=V'(\phi_1)=0,
\ee
The renormalization group improved potential has the form
\be
  \label{Veff}
  V(\phi) \propto \lambda(\phi)\phi^4 \left[
    1+O\left( \frac{\alpha}{4\pi}\log(M_i/M_j) \right)
  \right],
\ee
where $\alpha$ is  the common name for the SM coupling constants, and
$M_i$ are the masses of different particles in the background of the
Higgs field. So, instead of computing the effective potential, one can
solve the ``criticality equations'':
\be
\lambda(\mu_0)=0,\qquad\beta_\lambda^\text{SM}(\mu_0)=0.
\label{crit}
\ee
This simplified procedure works with accuracy better than $\delta y_t\simeq 0.001$ if $\lambda$ is taken in \MSb{} scheme.

In numbers, the criticality equations (\ref{crit}) give
\begin{multline}
  \label{ytcrit}
  y_t^{\text{crit}} = 0.9244 + 0.0012\times\frac{ M_h/{\rm GeV} -125.7}{0.4}\\
  +0.0012\times\frac{\alpha_s(M_Z)-0.1184}{0.0007},
\end{multline}
where  $\alpha_s$ is the QCD coupling at the $Z$-boson mass.  Though all the required components are present in the works \cite{Bezrukov:2012sa,Degrassi:2012ry,Buttazzo:2013uya,Pikelner:2014private} a comment is now in order of how eq.~(\ref{ytcrit}) was obtained. First, instead of defining the critical Higgs boson mass $M_h$ the critical value of the top pole mass was defined, and then converted back to the value of the top quark Yukawa, accounting for known QCD and electroweak corrections.  However, it is not immediate to read these numbers from the papers mentioned, as far as the matching conditions relating the physical masses and \MSb{} parameters are scattered over the published works.  The 3 loop beta functions can be found in \cite{Mihaila:2012fm,Mihaila:2012pz,Chetyrkin:2012rz,Chetyrkin:2013wya, Bednyakov:2012en,Bednyakov:2013eba} and is given in a concise form in the code from \cite{Bezrukov:2012sa} or in \cite{Buttazzo:2013uya}.  The one loop contributions to the matching conditions between the  $W$, $Z$ and Higgs boson masses and the \MSb{} coupling constants at $\mu\sim m_t$ of the order $O(\alpha)$ and $O(\alpha_s$) are known for long time \cite{Sirlin:1985ux} and can be read of \cite{Bezrukov:2012sa,Buttazzo:2013uya}.  The two loop contribution of the order $O(\alpha\alpha_s)$ to the Higgs coupling constant $\lambda$ was calculated in \cite{Bezrukov:2012sa,Buttazzo:2013uya} and for the practical purposes can be taken from eq.~(34) of \cite{Buttazzo:2013uya}.  The two loop contribution of the order $O(\alpha^2)$ to $\lambda$ was calculated in \cite{Buttazzo:2013uya}, with the numerical approximation given by eq.~(35).  Recently an independent evaluation at the order $O(\alpha^2)$ was obtained in \cite{Pikelner:2014private} which differs slightly from \cite{Buttazzo:2013uya}, but the difference has a completely negligible impact on (\ref{ytcrit}) (note that even the whole $O(\alpha^2)$ contribution to $\lambda$ changes $y_t^{\text{crit}}$ by only $0.5\times10^{-3}$).  However, one should be careful in using the final numerical values of the \MSb{} couplings from the section 3 of \cite{Buttazzo:2013uya}, as far as the value of the strong coupling at $\mu=M_t$ which was used there (eq.~(60)) does not correspond to the value obtained from the Particle Data Group value at $M_Z$ by RG evolution.

Thanks to complete two-loop computations of \cite{Buttazzo:2013uya,Pikelner:2014private} and three-loop beta functions for the SM couplings found in \cite{Mihaila:2012fm,Mihaila:2012pz,Chetyrkin:2012rz,Chetyrkin:2013wya,Bednyakov:2012en,Bednyakov:2013eba} the formula (\ref{ytcrit}) may have a very small theoretical error, $2\times10^{-4}$, with the latter number coming from an ``educated guess'' estimates of even higher order terms---4 loop beta functions for the SM and 3 loop matching conditions at the electroweak scale, which relate the physically measured parameters such as $W,~Z$ and Higgs boson masses, etc with the \MSb{} parameters (see the discussion in \cite{Bezrukov:2012sa} and more recently in \cite{Shaposhnikov:2013ira}). We stress that the experimental value of the mass of the top quark is \emph{not used in this computation}, we will come to this point later in Section~\ref{sec:comp}.

\begin{figure}
  \centering
  \includegraphics{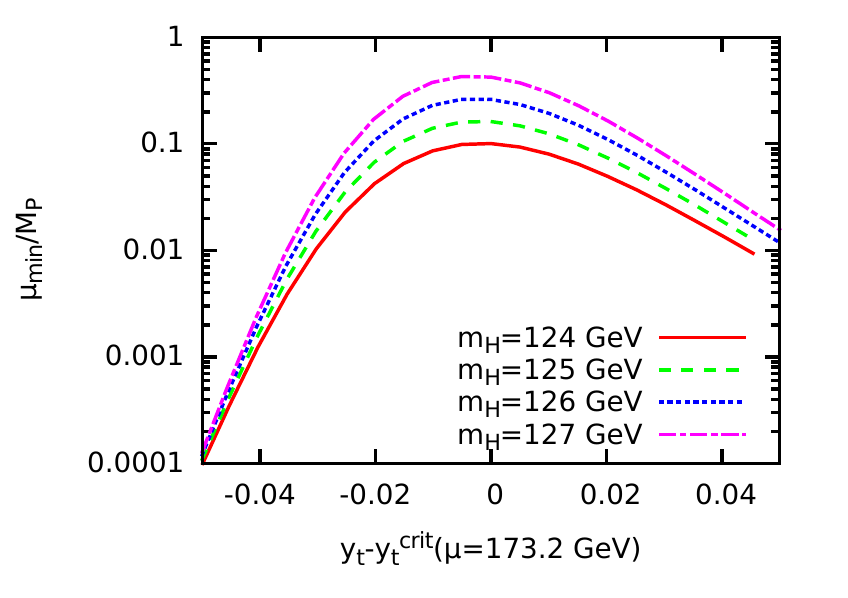}
  \caption{Scale of the minimum of the Higgs boson self-coupling depending on the top quark Yukawa $y_t(\mu=173.2\GeV)$ near the critical value $y_t^\text{crit}$.}
  \label{fig:mumin}
\end{figure}

Yet another interesting quantity which can be derived from eq.~(\ref{crit}) is the ``criticality'' scale $\mu_0$, where both the scalar self-coupling and its $\beta$-function are equal to zero.  Fig.~\ref{fig:mumin} contains its plot as a function of the top quark Yukawa for several Higgs masses.  It is amazing that $\mu_0$ happens to be very close to the reduced Planck scale $M_P$: taking the SM parameters as an input we get $\mu_0$ numerically very close to the  scale of gravity! This fact has been noted a long time ago in \cite{Froggatt:1995rt}  and may indicate the asymptotically safe character of the Standard Model and gravity, as has been discussed in \cite{Shaposhnikov:2009pv}. In the recent work \cite{Gorsky:2014una} it was argued that this may be a consequence of enhanced conformal symmetry at the Planck scale. At the same time, it could be a pure coincidence. It is also interesting to note that the extremum of $\mu_0$ as a function of the top quark Yukawa coupling (with other parameters fixed) is maximal at $y_t$ close to $y_t^\text{crit}$. We have no clue why this is so.

\section{Top Yukawa coupling and experiment}
The top Yukawa coupling can be extracted from a number of experiments. At present, the most precise determination of $y_t$ comes from the analysis of hadron collisions at Tevatron in Fermilab and LHC at CERN.  A specific parameter (called  Mont-Carlo (MC) top mass) in the event generators  such as PYTHIA \cite{Sjostrand:2007gs,Sjostrand:2006za} or HERWIG \cite{Bahr:2008pv}, is used to fit the data. The most recent determinations of MC top mass  are
\(
M_t = 173.34 \pm 0.27 \text{(stat)} \pm 0.71 \text{(syst)} \GeV
\)
from the combined analysis of ATLAS, CMS, CDF, and D0 (at 8.7~fb$^{-1}$ of Run II of Tevatron) \cite{ATLAS:2014wva},  \(
M_t = 174.34 \pm 0.37 \text{(stat)} \pm 0.52 \text{(syst)} \GeV
\)
from the CDF and D0 combined analysis of Run I and Run II of Tevatron \cite{Tevatron:2014cka},
and  \(
M_t = 172.38 \pm 0.10 \text{(stat)} \pm 0.65 \text{(syst)} \GeV
\)
from the CMS alone \cite{CMS:2014hta} (at 25~fb$^{-1}$ of Run I of LHC) .

The problem at hand is to compute the top quark Yukawa coupling in the \MSb{} scheme, which was used in the previous sections, from the MC top quark mass and other relevant electroweak parameters, determined experimentally. Unfortunately, there are no theoretical computations  relating these  quantities with the error bars small enough to make a clear cut determination of  the scale of new physics. Presumably, the best way to proceed would be to have an event generator where it is the top Yukawa coupling in the \MSb{} scheme\footnote{Or any other parameter that has a well-defined infrared safe relation to the Yukawa copling.} (rather than MC top mass), enters directly in the computation of different matrix elements.  Then the generated events can be  compared with the experimental one, leading to the direct determination of $y_t$.

At present the extraction of $y_t$ from experiment proceeds in a somewhat different way\footnote{The difficulties in extraction of $y_t$ from experiments at the LHC or Tevatron are discussed in \cite{Alekhin:2012py, Frixione:2014ala}. }. The analysis goes as follows. 

First, it is assumed that the MC top mass, taken from the analysis of  the decay products of the top quark, is close to the pole  mass, with the difference of the order of 1~GeV \cite{Buckley:2011ms,Juste:2013dsa,Frixione:2014ala}. Second, the pole top mass is related to the top Yukawa coupling, accounting for strong and electroweak corrections \cite{Bezrukov:2012sa,Buttazzo:2013uya}. 

Presently, the largest theoretical uncertainty is associated  with  the first step \cite{Frixione:2014ala}.  Yet another source of uncertainties may come from the fact, that, to the best of our knowledge, the electroweak effects are not included in MC generators \cite{Buckley:2011ms}.  This,  naively, could  introduce a relative error of the order of  ${\cal O}(\alpha_W/\pi)\sim 10^{-2}$ in the pole mass of the top quark. 

The second step adds further ambiguities. The pole quark mass is not well defined theoretically, since the top quark carries colour and thus does not exist as an asymptotic state. The  non-perturbative QCD effects of the order of $ \Lambda_{QCD} \simeq \pm 300$~MeV would lead to $\delta y_t/y_t \sim 10^{-3}$. The similar in amplitude effect comes from (unknown) ${\cal O}(\alpha_s^4)$ corrections to the relation between the pole and $\rm{\overline{MS}}$ top quark masses. According to \cite{Kataev:2010zh}, this correction can be as large as  $\delta y_t/y_t \simeq -750 (\alpha_s/\pi)^4\simeq - 0.002$. 

The theoretically more clean extraction of the top Yukawa coupling comes from the measurements of the total cross-section of the top production \cite{Alekhin:2012py} that can be directly calculated in the \MSb{} scheme, but it has much larger errors.

\begin{figure*}
  \centering
  \includegraphics{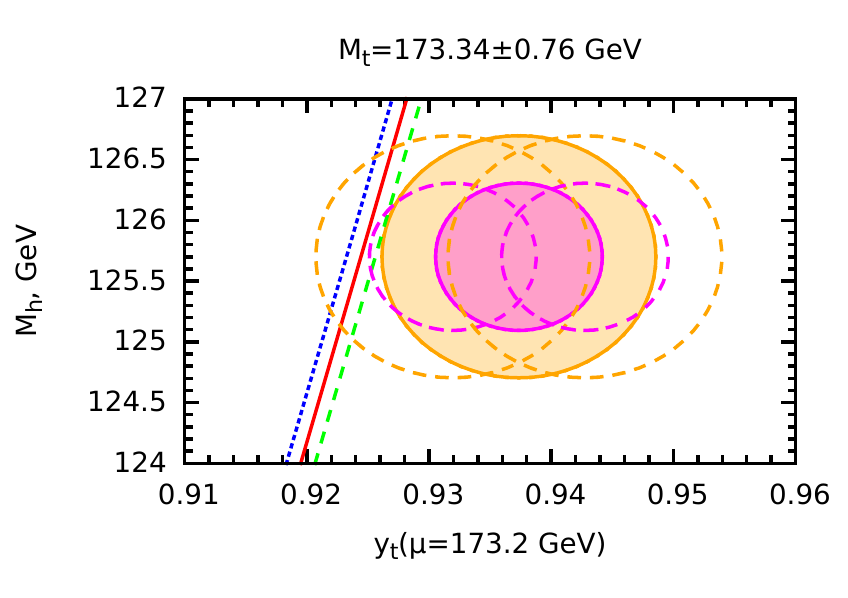}%
  \includegraphics[width=3.5in]{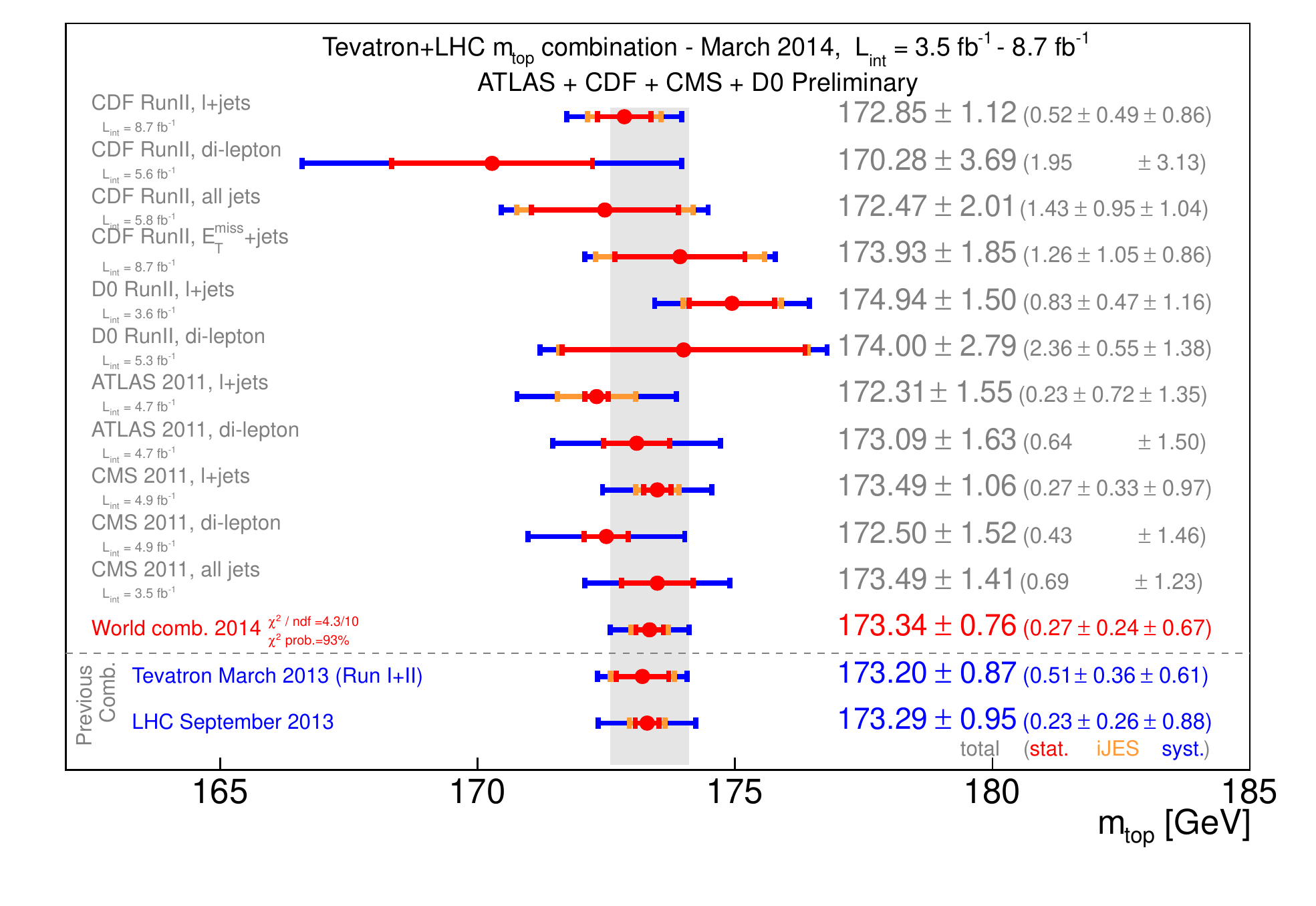}
  \caption{The plot demonstrating the relation of the current measurements of the top quark mass $M_t$ and the critical value of the top quark Yukawa $y_t$.  The diagonal line is the critical value of the Yukawa coupling, with the uncertainties associated with the experimental error of the $\alpha_s$ indicated by dashed lines.  To the left of these lines the SM vacuum is absolutely stable and to the right it is metastable. The filled ellipses correspond to the 1 and $2\sigma$ \emph{experimental} errors of the determination of the top quark MC mass, converted to the Yukawa top using as if it were the  pole mass.  Dashed ellipses demonstrate the possible shift due to ambiguous relation of the pole and MC masses.  The top quark mass is from the combined LHC and Tevatron analysis
\cite{ATLAS:2014wva}, with the individual experiments results are shown on the right (plot from \cite{ATLAS:2014wva}).}
  \label{fig:mhyt}
\end{figure*}

\begin{figure*}
  \centering
  \includegraphics{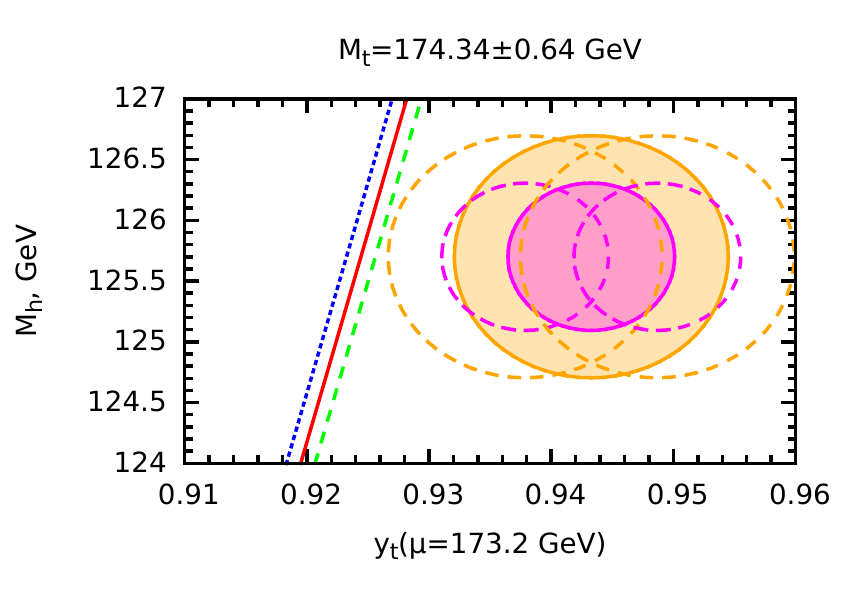}%
  \includegraphics[width=2.5in]{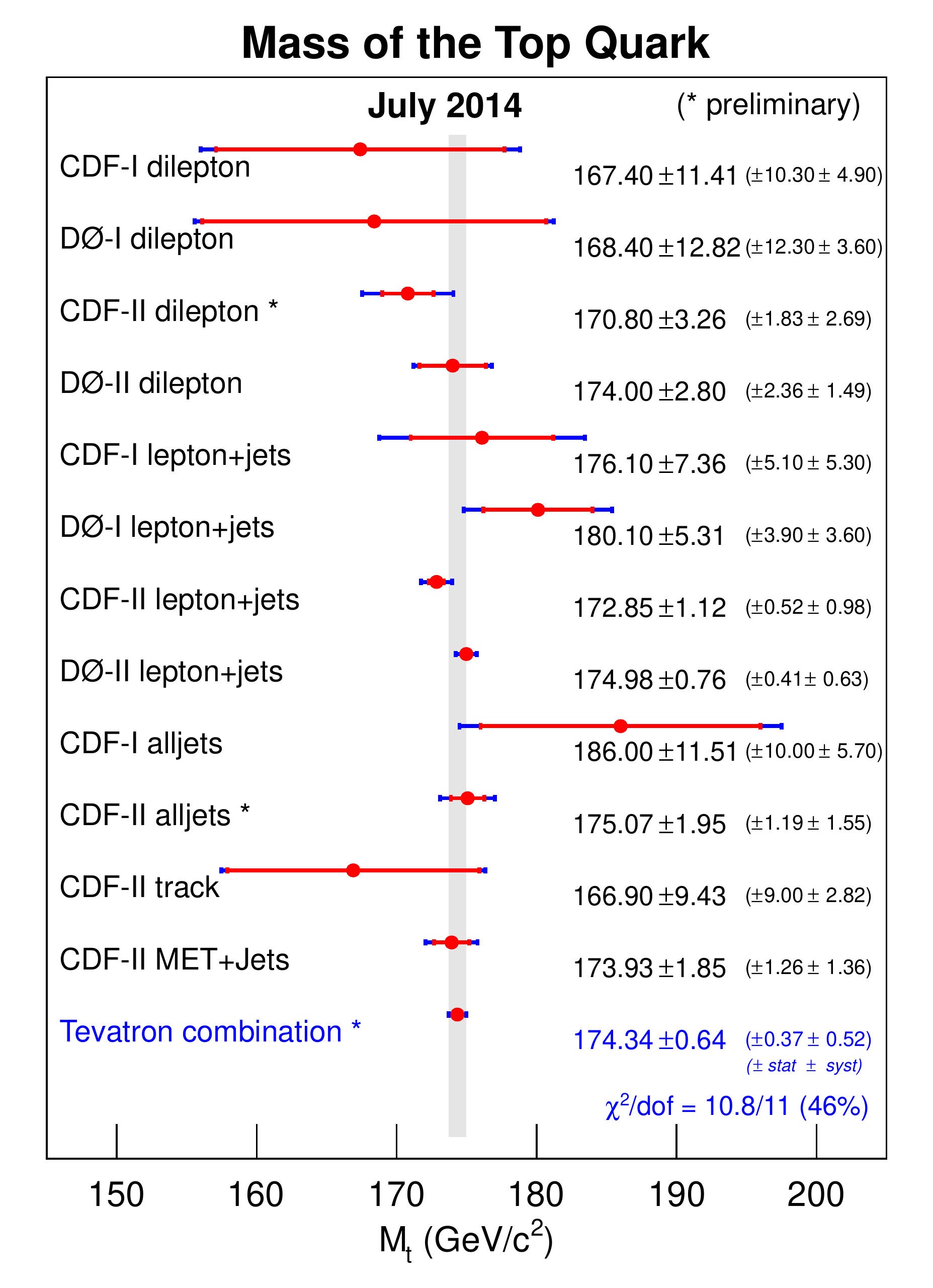}
  \caption{The same as Fig.~\ref{fig:mhyt} but for the higher top mass reported by the latest Tevatron analysis \cite{Tevatron:2014cka}.  Right plot (taken from \cite{Tevatron:2014cka}) indicates the individual measurements.}
\label{fig:mhyt-tevatron}
\end{figure*}

\begin{figure*}
  \centering
  \includegraphics{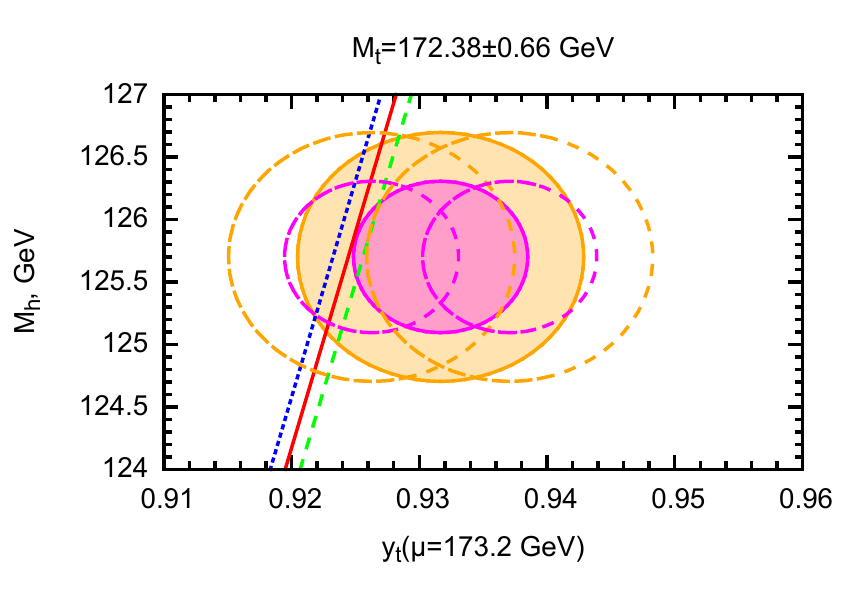}%
  \includegraphics[width=2.5in]{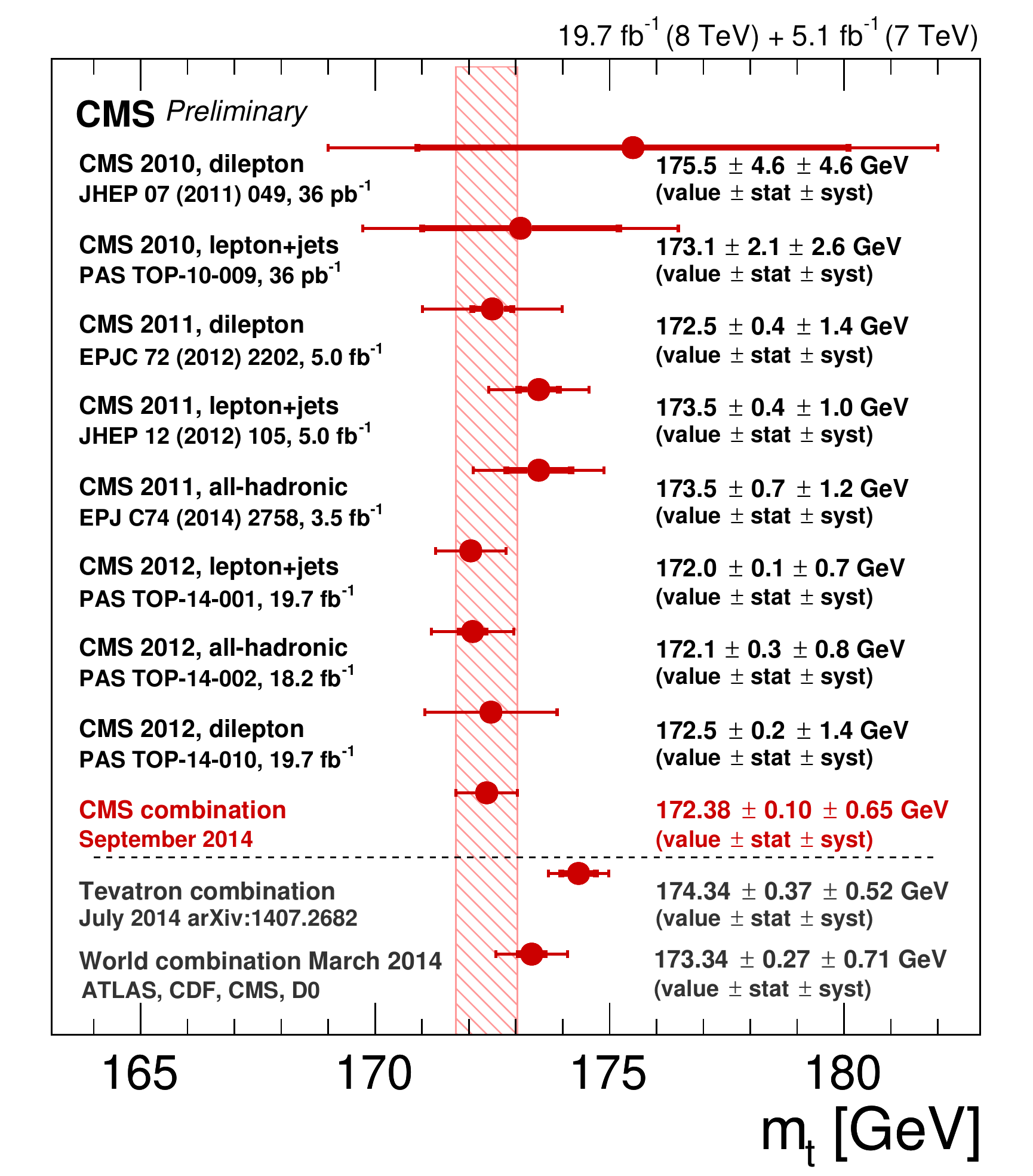}
  \caption{The same as Fig.~\ref{fig:mhyt} but for the \emph{lower} top mass reported by CMS from LHC run I \cite{CMS:2014hta}.  Right plot (taken from \cite{CMS:2014hta}) indicates the individual measurements.}
\label{fig:mhyt-cms}
\end{figure*}

In Figs.~\ref{fig:mhyt},~\ref{fig:mhyt-tevatron},~\ref{fig:mhyt-cms} we show the comparison between experiment and the theoretical computation of the critical value of the top Yukawa  coupling. The difference between the two is within 1--3 standard deviations, accounting for systematic uncertainties. In other words, it is perfectly possible that our vacuum is absolutely stable and the SM  is a valid theory up to the Planck scale even in the cosmological context. It is also perfectly possible that it is another way around and that we need some kind of  new physics at energies around  $10^7$~GeV or below. 

\section{Conclusions}
Obviously, the energy scale of new physics is  crucial for the possible outcome of the high energy (LHC \cite{LHChttp}, FCC \cite{FCChttp}, ILC \cite{ILChttp}), intensity (LHCb \cite{bLHChttp}, SHiP \cite{SHIPhttp}) and accuracy (searches for baryon and lepton number violation, LAGUNA \cite{LAGUNAhttp}, LBNE \cite{LBNEhttp}) frontiers of high energy physics. The theoretical prejudice about the scale of new physics is quite subjective and does not give a unique answer, especially given the discovery of the Higgs boson with a very peculiar value of its mass and the absence of deviations from the Standard Model in accelerator experiments. Under these circumstances  the precise measurement of the top quark Yukawa coupling is very important.

Variation of the top quark Yukawa coupling in the allowed by experimental and theoretical uncertainties interval changes the place where the scalar self-coupling crosses zero from $10^7$~GeV to infinity, without a clear indication of the necessity of new thresholds in particle physics between the Fermi and Planck scales. For the largest allowed top Yukawa coupling (we take 2 sigma in determination of the Monte-Carlo top mass and add to it $1$~GeV uncertainty in comparison between the pole and MC masses) the scale  $\mu_\text{new}$ is as small as $10^7$~GeV, whereas if the uncertainties are pushed in the other direction no new physics would be needed below the Planck mass.

A precise measurement of $y_t$ would be possible at $e^+e^-$ colliders such as ILC \cite{ILChttp} or FCC-ee \cite{TLEPhttp}. Otherwise, a theoretical breakthrough in understanding of the precise top Yukawa extraction from pp collisions is needed.  At present, the evidence for new physics beyond the SM coming from the top and Higgs mass measurements is at the level of 1--3$\sigma$, having roughly the same statistical significance as other reported anomalies, for example muon magnetic moment \cite{Bennett:2006fi}, MiniBooNE \cite{AguilarArevalo:2008rc} and LSND \cite{Aguilar:2001ty}.  It remains to be seen which of them (if any) will eventually be converted into undisputed signal of new physics between the Fermi and Planck scales.

\bigskip

The work of M.S. is supported in part by the Swiss National Science Foundation.
F.B. would like to thank CERN, where this paper was writen, for hospitality.
We would like to thank  Abdelhak Djouadi, Stefano Frixione, and Andrey Pikelner for many helpful discussions related to this paper.

\bibliographystyle{apsrev4-1-fix}
\bibliography{local}

\end{document}